\documentclass[a4paper,UKenglish,cleveref,autoref,thm-restate]{oasics-v2019}
\nolinenumbers

\newcommand{\SA}{\ensuremath{\mathrm{SA}}}
\newcommand{\rank}{\ensuremath{\mathrm{rank}}}
\newcommand{\select}{\ensuremath{\mathrm{select}}}
\newcommand{\LF}{\ensuremath{\mathrm{LF}}}
\newcommand{\rev}{\ensuremath{\mathrm{rev}}}

\title{Faster run-length compressed suffix arrays}

\titlerunning{Faster RLCSAs}

\author{Nathaniel K.\ Brown}{Department of Computer Science, Johns Hopkins University, USA}{nbrown99@jh.edu}{https://orcid.org/0000-0002-6201-2301}{Supported in part by a Johns Hopkins University Computer Science PhD Fellowship and NIH grants R21HG013433 and R01HG011392.}

\author{Travis Gagie\footnote{Corresponding author.}}{Faculty of Computer Science, Dalhousie University, Canada}{travis.gagie@dal.ca}{https://orcid.org/0000-0003-3689-327X}{Funded in part by NSERC grant RGPIN-07185-2020.}

\author{Giovanni Manzini}{Department of Computer Science, University of Pisa, Italy}{giovanni.manzini@unipi.it}{https://orcid.org/0000-0002-5047-0196}{Partially supported by the NextGeneration EU programme PNRR ECS00000017 Tuscany Health Ecosystem (Spoke 6, CUP: I53C22000780001) and by the project PAN-HUB funded by the Italian Ministry of Health (ID: T4-AN-07, CUP: I53C22001300001)}

\author{Gonzalo Navarro}{Department of Computer Science, University of Chile, Chile}{gnavarro@dcc.uchile.cl}{https://orcid.org/0000-0002-2286-741X}{Funded in part by CeBiB under Basal Funds FB0001 and AFB240001, ANID, Chile.}

\author{Marinella Sciortino}{Department of Mathematics and Computer Science, University of Palermo, Italy}{marinella.sciortino@unipa.it}{https://orcid.org/0000-0001-6928-0168}{Partially supported by the project ``ACoMPA'' (CUP B73C24001050001) funded by the NextGeneration EU programme PNRR ECS00000017 Tuscany Health Ecosystem (Spoke 6) and by MUR PRIN project ``PINC'' (no. 2022YRB97K)}

\authorrunning{N.K.\ Brown, T.\ Gagie, G.\ Manzini, G.\ Navarro and M.\ Sciortino}

\Copyright{Nathaniel K.\ Brown, Travis Gagie, Giovanni Manzini, Gonzalo Navarro and Marinella Sciortino}



\acknowledgements{Many thanks to the first author's CSCI 4419 / 6106 students Anas Alhadi, Nour Allam, Dove Begleiter, Nithin Bharathi {Kabilan Karpagavalli}, Suchith {Sridhar Khajjayam} and Hamza Wahed, and to Christina Boucher, Ben Langmead, Jouni Sir\'en and Mohsen Zakeri.}

\EventEditors{John Q. Open and Joan R. Access}
\EventNoEds{2}
\EventLongTitle{42nd Conference on Very Important Topics (CVIT 2016)}
\EventShortTitle{CVIT 2016}
\EventAcronym{CVIT}
\EventYear{2016}
\EventDate{December 24--27, 2016}
\EventLocation{Little Whinging, United Kingdom}
\EventLogo{}
\SeriesVolume{42}
\ArticleNo{23}

\begin{document}

\maketitle

\begin{abstract}
We first review how we can store a run-length compressed suffix array (RLCSA) for a text $T$ of length $n$ over an alphabet of size $\sigma$ whose Burrows-Wheeler Transform (BWT) consists of $r$ runs in $O \left( \rule{0ex}{2ex} r \log (n / r) + r \log \sigma + \sigma \right)$ bits such that later, given character $a$ and the suffix-array (SA) interval for $P$, we can find the SA interval for $a P$ in $O (\log r_a + \log \log n)$ time, where $r_a$ is the number of runs of copies of $a$ in the BWT.  We then show how to modify the RLCSA such that we find the SA interval for $a P$ in only $O (\log r_a)$ time, without increasing its asymptotic space bound.  Our key idea is applying a result by Nishimoto and Tabei (ICALP 2021) and then replacing rank queries on sparse bitvectors by a constant number of select queries.  We also review two-level indexing and discuss how our faster RLCSA may be useful in improving it.  Finally, we briefly discuss how two-level indexing may speed up a recent heuristic for finding maximal exact matches of a pattern with respect to an indexed text.
\end{abstract}

\section{Introduction}
\label{sec:introduction}

Grossi and Vitter's compressed suffix arrays (CSAs)~\cite{GV05} and Ferragina and Manzini's FM-indexes~\cite{FM05} are sometimes treated as almost interchangeable, but their query-time bounds are quite different.  With a CSA for a text $T$ of length $n$ over an alphabet of size $\sigma$, when given a character $a$ and the suffix-array (SA) interval for a pattern $P$ we can find the SA interval for $a P$ in $O (\log n_a)$ time, where $n_a$ is the number of occurrences of $a$ in the text; with an FM-index we use $O (\log \sigma)$ time.  This difference carries over to run-length compressed suffix arrays (RLCSAs)~\cite{MNSV10,Sir12} and run-length compressed FM-indexes (RLFM-indexes)~\cite{GNP20,MN05}, with both taking space proportional to the number $r$ of runs in the Burrows-Wheeler Transform (BWT) of the text but the former being generally faster for texts over large alphabets with relatively few runs of each character, and the latter being faster for texts over smaller alphabets.

In Section~\ref{sec:review} we review (with some artistic license) CSAs, RLCSAs and interpolative coding and show how an RLCSA for $T$ takes $O \left( r \log (n / r) + r \log \sigma + \sigma \right)$ bits and allows us to find the SA interval for $a P$ from that of $P$ in $O (\log r_a + \log \log n)$ time, where $r_a$ is the number of those runs in the BWT containing copies of $a$.  In Section~\ref{sec:NT21} we review a result by Nishimoto and Tabei~\cite{NT21} about splitting the runs in the BWT so that we can evaluate $\LF$ in constant time, without increasing the number of runs by more than a constant factor.  In Section~\ref{sec:speedup} we present our main result: how to modify the RLCSA from Section~\ref{sec:review} such that finding the SA interval for $a P$ takes only $O (\log r_a)$ time, without increasing the asymptotic space bound.  In Section~\ref{sec:two-level_indexing} we discuss two-level indexing, for which we build one index for the text and another for the parse of the text, and how our faster RLCSA may be more suitable for indexing parses than current options.  Finally, in Section~\ref{sec:MEM-finding} we discuss a related heuristic for finding longest maximal exact matches (MEMs) of a pattern with respect to an indexed text.

\section{Preliminaries}
\label{sec:review}

\subsection{Compressed suffix arrays}

The key idea behind compressed suffix arrays (CSAs) is to store $\Psi [0..n - 1]$ compactly while supporting certain searches on it quickly, where $\Psi [0..n - 1]$ is the permutation of $\{0, \ldots, n - 1\}$ such that $\Psi [i]$ is the position of suffix-array entry $(\SA [i] + 1) \bmod n$ in $\SA [0..n - 1]$ or, equivalently, the position in $L$ of $F [i]$.  (This means $\Psi$ is the inverse of the $\LF$ mapping used in FM-indexes.)  By the definition, $\Psi$ consists of at most $\sigma$ increasing intervals --- one for each distinct character that occurs in the text, corresponding to the interval of suffixes starting with that character --- and if we can support fast binary searches on these intervals then we can support fast pattern matching.

\begin{figure}[p!]
\begin{center}
\includegraphics[width=0.9\textwidth]{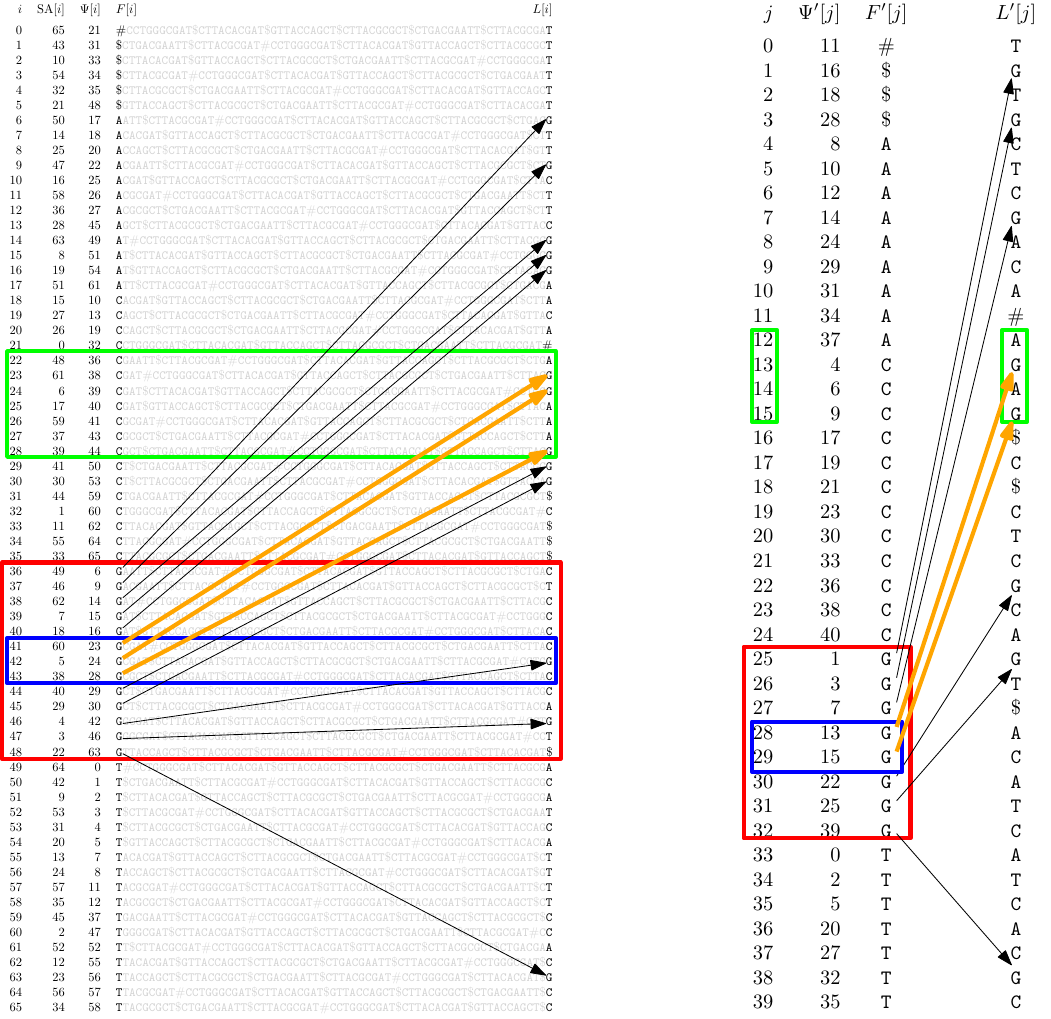}
\caption{For
\[T = \mathtt{CCTGGGCGAT\$CTTACACGAT\$GTTACCAGCT\$CTTACGCGCT\$CTGACGAATT\$CTTACGCGAT\#}\] we show $\SA$, $\Psi$, $F$ and $L$ on the left and the $\Psi'$, $F'$ and $L'$ on the right.  If we know $\SA [22..28]$ is the SA interval for {\tt CG} (in the green rectangle on the left) and we want the SA interval for {\tt GCG}, then we can search in the increasing interval
\[\Psi [36..48] = 6, 9, 14, 15, 16, 23, 24, 28, 29, 30, 42, 46, 63\]
for {\tt G} (in the red rectangle on the left, with $\Psi$ values between 22 and 28 shown as orange arrows and the others shown as black arrows) for the successor $\Psi [41] = 23$ of 22 and the predecessor $\Psi [43] = 28$ of 28.  We thus learn that the SA interval for {\tt GCG} is $\SA [41..43]$ (in the blue rectangle on the left).\\[1ex]
On the other hand, if we know $\SA [22..28]$ starts at offset 0 in the $L$ run of character $L' [12]$ --- that is, at offset 0 in the 13th run, counting from 1 --- and ends at offset 1 in the $L$ run of character $L' [15]$ (in the green rectangle on the right), then we can search in the increasing interval
\[\Psi' [25..32] = 1, 3, 7, 13, 15, 22, 25, 39\]
for {\tt G} (in the red rectangle, with $\Psi'$ values between 12 and 15 shown as orange arrows and the others shown as black arrows) for the successor $\Psi' [29] = 13$ of 12 and the predecessor $\Psi' [30] = 15$ of 15 (in the blue rectangle on the right).  We then use select and rank queries on two $n$-bit sparse vectors to find the SA interval for {\tt GCG}, the $L$ runs containing that interval's starting and ending positions, and those positions' offsets in those runs.}
\label{fig:CSAs}
\end{center}
\end{figure}

For example, consider the text \[T = \mathtt{CCTGGGCGAT\$CTTACACGAT\$GTTACCAGCT\$CTTACGCGCT\$CTGACGAATT\$CTTACGCGAT\#}\,,\]
for which $\SA$, $\Psi$, $F$ and $L$ are shown on the left in Figure~\ref{fig:CSAs}.  If we know $\SA [22..28]$ is the SA interval for {\tt CG} (in the green rectangle) and we want the SA interval for {\tt GCG}, then we can search in the increasing interval
\[\Psi [36..48] = 6, 9, 14, 15, 16, 23, 24, 28, 29, 30, 42, 46, 63\]
for {\tt G} (in the red rectangle, with $\Psi$ values between 22 and 28 shown as orange arrows and the others shown as black arrows) for the successor $\Psi [41] = 23$ of 22 and the predecessor $\Psi [43] = 28$ of 28.  We thus learn that the SA interval for {\tt GCG} is $\SA [41..43]$ (in the blue rectangle).  Knowing this, we can continue backward stepping.

\subsection{Run-length compressed suffix arrays revisited}

Run-length compressed suffix array (RLCSA) were introduced in~\cite{Sir12} for indexing highly repetitive collections. In this section we present an alternative, but functionally equivalent, description of RLCSAs which is more suitable for describing our improvements.

\begin{definition}
For a text $T [0..n - 1]$, the array $L' [0..r - 1]$ stores the sequence of $r$ characters in the runs in the run-length encoding of $L$.
\end{definition}

\begin{definition}
For a text $T [0..n - 1]$, the array $F' [0..r - 1]$ stores the $r$ characters in $L'$ in lexicographic order.
\end{definition}

\begin{definition}
For a text $T [0..n - 1]$, the array $\Psi' [0..r - 1]$ is the permutation of $\{0, \ldots, r - 1\}$ such that $\Psi' [i]$ is the position $F' [i]$ in $L'$.
\end{definition}

\noindent In this paper we view a RLCSA as a data structure storing $\Psi' [0..r - 1]$ compactly while supporting certain searches on it quickly.
By the definition of $\Psi'$, it still consists of at most $\sigma$ increasing intervals --- one for each distinct character that occurs in $T$, corresponding to the interval of suffixes starting with that character --- and if we can still support fast binary searches on these intervals then we can still support fast pattern matching.

For example, consider \[T = \mathtt{CCTGGGCGAT\$CTTACACGAT\$GTTACCAGCT\$CTTACGCGCT\$CTGACGAATT\$CTTACGCGAT\#}\]
again, for which $\Psi'$, $F'$ and $L'$ are shown on the right in Figure~\ref{fig:CSAs}.  If we know the SA interval $\SA [22..28]$ for {\tt CG} starts at offset 0 in the $L$ run of character $L' [12]$ and ends at offset 1 in the $L$ run of character $L' [15]$ (in the green rectangle) and we want the SA interval for {\tt GCG}, then we can search in the increasing interval
\[\Psi' [25..33] = 1, 3, 7, 13, 15, 22, 25, 39\]
for {\tt G} (in the red rectangle, with $\Psi'$ values between 12 and 15 shown as orange arrows and the others shown as black arrows) for the successor $\Psi' [28] = 13$ of 12 and the predecessor $\Psi' [29] = 15$ of 15.

Because $L'$ and $F'$ do not have the predecessor-successor relationship of $L$ and $F$, we cannot deduce that the SA interval for {\tt GCG} starts in the $L$ run of character $L' [28]$ and ends in the $L$ run of character $L' [29]$ (and, in fact, in this example it does not).  Instead, we store two $n$-bit SD-bitvectors~\cite{OS07}, $B_L$ and $B_F$, with $r$ copies of $1$ each.  The 1s in $B_L$ mark the starting positions of runs in $L$ and the 1s in $B_F$ mark the positions in $F$ of the marked characters in $L$.  In our example
\begin{eqnarray*}
B_F & = & \mathtt{11100110111001111111111100010111011011100\,\textcolor{blue}{\bf 101}\,0011110000010101111000} \\ 
B_L & = & \mathtt{10000011011101100101111101001001110011100\,\textcolor{blue}{\bf 011}\,0111111111110001011110}\,.
\end{eqnarray*}

\noindent The interval $B_F [41..43]$ in $B_F$ starting immediately before the bit with offset 0 in the block whose starting position is marked with the 29th copy of 1 and ending immediately before the bit with offset 1 in the block whose starting position is marked with the 30th copy of 1, is shown in blue.  (We are interested in the blocks marked with the 29th and 30th copies of 1 because we count from 0 in the $j$ column in Figure~\ref{fig:CSAs}, so those blocks correspond to $\Psi' [28]$ and $\Psi' [29]$.)  We can find this interval with 2 $\select_1$ queries on $B_F$, which take constant time.

The corresponding interval $B_L [41..43]$ in $B_L$ is also shown in blue, starting immediately before the bit with offset 3 in the block whose starting position is marked with the 22nd copy of 1 and ending immediately before the bit with offset 1 in the block whose starting position is marked with the 24th copy of 1.  We can find the 2 indices 22 and 24 with 2 $\rank_1$ queries on $B_L$, which take $O (\log \log n)$ time.  This means the SA interval for {\tt GCG} is $\SA [41..43]$ and it starts at offset 3 in the $L$ run of character $L' [21]$ and ends at offset 1 in the $L$ run of character $L' [23]$.  Knowing this we can continue backward stepping.

The RLCSA in Sir\'en's PhD thesis~\cite{Sir12} for a text $T[0..n - 1]$ with $r$ BWT runs takes $O \left( \rule{0ex}{2ex} r \log (n / r) + r \log \sigma + \sigma \log n \right)$ bits. Given a character $a$ and the SA interval for $P$, it can find the SA interval for $a P$ in $O (\log n)$ time.

\section{Faster RLCSAs}

\subsection{Searchable Interpolative coding}

Suppose we are given an increasing list $\ell_1, \ldots, \ell_k$ of $k$ integers in the range $[0..n - 1]$.  To encode them with {\em interpolative coding}~\cite{MS00}, we first write $\ell_{\lceil k / 2 \rceil}$ using $\lfloor \lg (n - 1) \rfloor + 1$ bits (except that we write 0 using 1 bit).  All the numbers $\ell_1, \ldots, \ell_{\lceil k / 2 \rceil - 1}$ are in the range $[0..\ell_{\lceil k / 2 \rceil} - 1]$, so we can encode them recursively.  All the numbers $\ell_{\lceil k / 2 \rceil + 1}, \ldots, \ell_k$ are in the range $[\ell_{\lceil k / 2 \rceil} + 1..n - 1]$, so we can encode them recursively as $\ell_{\lceil k / 2 \rceil + 1} - \ell_{\lceil k / 2 \rceil} - 1, \ldots, \ell_k - \ell_{\lceil k / 2 \rceil} - 1$.  Each encoding has $O (\log n)$ bits, so we can read them in $O (1)$ time.

For example, if $n = 66$, $k = 13$ and the list is $6, 9, 14, 15, 16, 23, 24, 28, 29, 30, 42, 46, 63$, then we start by encoding $\ell_7 = 24$ using $\lfloor \lg 65 \rfloor + 1 = 7$ bits as {\tt 0011000}.  We then encode $\ell_3 = 14$ using $\lfloor \lg 23 \rfloor + 1 = 5$ bits as {\tt 01110}.  We then encode $\ell_1, \ell_2, \ell_5, \ell_4, \ell_6 = 6, 9, 16, 15, 23$ as $\mathtt{0110}, \mathtt{010}, \mathtt{0001}, \mathtt{0}, \mathtt{110}$, and $\ell_{10}, \ell_8, \ell_9, \ell_{12}, \ell_{11}, \ell_{13}$ as $\mathtt{000101}, \mathtt{011}, \mathtt{0}, \mathtt{001111}, \mathtt{1011}, \mathtt{10000}$.

\begin{figure}[t!]
\begin{center}
\includegraphics[width=0.9\textwidth]{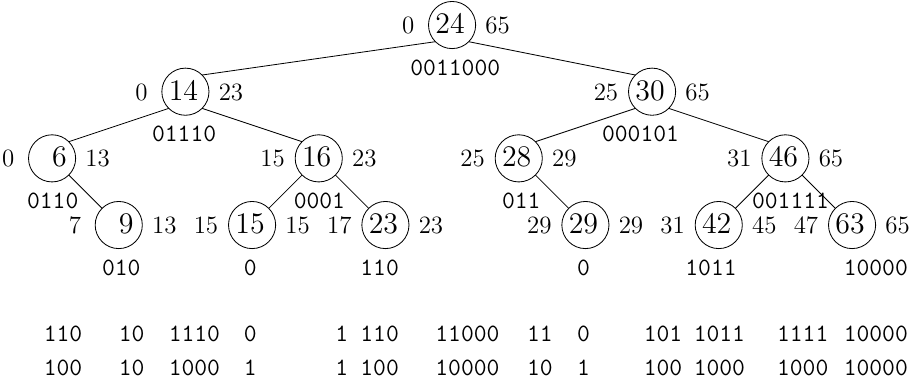}
\caption{A balanced binary search tree storing the $k = 13$ keys from the increasing list $6, 9, 14, 15, 16, 23, 24, 28, 29, 30, 42, 46, 63$ with each key in the range $[0..n - 1 = 65]$.  When we reach each key in a pre-order traversal or binary search, we know it lies between the two values show to its left and right, so we can encode it as the binary number shown below it, using a total of $O (k \log (n / k) + k)$ bits.  If we store a bitvector marking the start of each encoding as visited in an in-order traversal, as shown below the tree, then we can omit the leading 0s from the encodings and support binary search in time $O (\log k)$ without changing our asymptotic space bound.}
\label{fig:interpolative}
\end{center}
\end{figure}

If we imagine the list stored as keys in a balanced binary search tree, as illustrated in Figure~\ref{fig:interpolative}, we encode the keys according to a pre-order traversal: when we reach each key $\ell_i$, we know $\ell_i$ lies between the numbers shown to the left and right of $\ell_i$ and we encode $\ell_i$ using the maximum number of bits we would need for any key in that range.  For example, when we reach 46 in a pre-order traversal of the tree in Figure~\ref{fig:interpolative}, we know it lies between 31 and 65, so we encode it using $\lfloor \lg (65 - 31) \rfloor + 1 = 6$ bits as $(46 - 31)_2 = \mathtt{001111}$.

The binary search tree has height $\lfloor \lg k \rfloor$ and the bottom level contains at most $k$ keys.  By Jensen's Inequality, we encode those keys using $O (k \log (n / k) + k)$ bits.  Similarly, there at most $k / 2^h$ keys at height $h$ and we encode those keys using $O \left( \frac{k}{2^h} \log \frac{n}{k / 2^h} + \frac{k}{2^h} \right) = O \left( \frac{k}{2^h} \log (n / k) + \frac{k (h + 1)}{2^h} \right)$ bits.  Since
\[\sum_{h = 0}^{\lfloor \lg k \rfloor} \frac{k (h + 1)}{2^h} = O (k)\,,\]
we use $O (k \log (n / k) + k)$ bits in total.

In this paper we want to perform binary search on the list --- the reader may have noticed that our example is $\Psi [36..48]$ from Figure~\ref{fig:CSAs} --- so we want random access to the encodings of the numbers in it.  We can store the encodings according to an in-order traversal instead of a pre-order traversal, and store an uncompressed bitvector with as many bits as there are in the concatenation of the encodings and 1s marking where the encodings start.  Since the bitvector delimits the encodings, however, we can delete the leading 0s from each encoding before concatenating them and building the bitvector.  The in-order encodings for our example are shown below the tree in Figure~\ref{fig:interpolative}, with the leading 0s removed, and the bitvector is shown below them.  Since the bitvector uses at most as many bits as the encodings, we still use $O (k \log (n / k) + k)$ bits in total and we can perform binary search in $O (\log k)$ time.  This scheme is similar to Teuhola's~\cite{Teu11} and Claude, Nicholson and Seco's~\cite{CNS14}.

To find the successor of 22 in the list, we start at the root knowing $n = 66$ and $k = 13$ and perform $\select_1 (7)$ and $\select_1 (8)$ queries on the bitvector to find the starting and ending positions of the encoding {\tt 0011000} of $\ell_7 = 24$ in the range $[0..65]$.  Since $22 < 24$, we then perform $\select_1 (3)$ and $\select_1 (4)$ queries to find the starting and ending positions of the encoding {\tt 01110} of $\ell_3 = 14$ in the range $[0..23]$.  Since $22 > 14$, we then perform $\select_1 (5)$ and $\select_1 (6)$ queries to find the starting and ending positions of the encoding {\tt 0001} of $\ell_5 = 16$ in the range $[15..23]$.  Since $22 > 16$, we then perform $\select_1 (6)$ and $\select_1 (7)$ queries to find the starting and ending positions of the encoding {\tt 110} of $\ell_6 = 23$ in the range $[17..23]$.  Since $22 < 23$, we know the successor of 22 in $L$ is 23.  We can find the predecessor of 28 in $O (\log k)$ time symmetrically.

If we apply interpolative coding with fast binary search to the increasing interval of $\Psi$ for a character $a$ in a text $T$ of length $n$, then we use $O (n_a \log (n / n_a) + n_a)$ bits and can support binary search in $O (\log n_a)$ time, where $n_a$ is the frequency of $a$ in $T$.  If we do this for all the characters then we use $O (n (H_0 (T) + 1))$ bits, where $H$ is the 0th-order empirical entropy of $T$.  If we encode the increasing interval of $\Psi'$ for $a$ with interpolative coding, then we use $O (r \log (r / r_a) + r_a)$ bits and can support binary search in $O (\log r_a)$ time, where $r_a$ is the number of runs of copies of $a$ in the BWT of $T$ (and, equivalently, in $L$).  If we do this for all the characters then we use $O (r (H_0 (L') + 1))$ bits, where $L'$ is again the sequence of $r$ characters in the runs in the run-length encoding of $T$.  To be able to find the increasing interval for $a$ in $\Psi'$, we store an $r$-bit uncompressed bitvector with 1s marking the where the intervals start.

\begin{theorem}
\label{thm:interpolative}
We can store $\Psi'$ for $T$ in $O (r (H_0 (L') + 1)) \subseteq O (r \log \sigma)$ bits and support binary search in the increasing interval for a character $a$ in $O (\log r_a)$ time, where $r_a$ is the number of runs of copies of $a$ in the BWT of $T$.
\end{theorem}

\noindent Using Theorem~\ref{thm:interpolative} in an RLCSA gives us the following corollary, with the $O (\log \log n)$ term in the query-time bound coming only from the $O (\log \log n)$ query-time bound for rank on SD-bitvectors.  The $O (r \log (n / r))$ term in the space bound comes from the SD-bitvectors, the $O (r \log \sigma)$ term comes from the interpolative coding (and includes an $r$-bit bitvector marking where the increasing intervals for the distinct characters start in $F'$) and the $O (\sigma)$ term comes from a $\sigma$-bit uncompressed bitvector marking which distinct characters occur in $T$.

\begin{corollary}
\label{cor:interpolative}
We can store an RLCSA for $T$ in $O \left( \rule{0ex}{2ex} r \log (n / r) + r \log \sigma + \sigma \right)$ bits such that, given character $a$ and the SA interval for $P$, we can find the SA interval for $a P$ in $O (\log r_a + \log \log n)$ time.
\end{corollary}

\subsection{Splitting Theorem for RLCSAs}
\label{sec:NT21}

Nishimoto and Tabei~\cite{NT21} showed how we can split the runs in $L$ such that no block in $B_F$ overlaps more than a constant number of blocks in $B_L$ without increasing the number of runs by more than a constant factor, and then store $\LF$ in $O (r \log n)$ bits and evaluate it in constant time.  Brown, Gagie and Rossi~\cite{BGR22} slightly generalized their key theorem:

\begin{theorem}[Nishimoto and Tabei~\cite{NT21}; Brown, Gagie and Rossi~\cite{BGR22}]
\label{thm:NT21}
Let $\pi$ be a permutation on $\{0, \ldots, n - 1\}$,
$$P = \{0\} \cup \{i\ :\ 0 < i \leq n - 1, \pi (i) \neq \pi (i - 1) + 1\}\,,$$
and $Q = \{\pi (i)\ :\ i \in P\}$.  For any integer $d \geq 2$, we can construct $P'$ with $P \subseteq P' \subseteq \{0, \ldots, n - 1\}$ and $Q' = \{\pi (i)\ :\ i \in P'\}$ such that
\begin{itemize}
    \item if $q, q' \in Q'$ and $q$ is the predecessor of $q'$ in $Q'$, then $|[q, q') \cap P'| < 2 d$,
    \item $|P'| \leq \frac{d |P|}{d - 1}$.
\end{itemize}
\end{theorem}

\noindent If $L [i] = L [i - 1]$ then $\LF (i) = \LF (i - 1) + 1$, so
\[\{0\} \cup \{i\ :\ 0 < i \leq n - 1, \LF (i) \neq \LF (i - 1) + 1\}\]
has cardinality $r$.  If $\LF (i) = \LF (i - 1) + 1$ then, since $\Psi$ and $\LF$ are inverse permutations, $\Psi [j] = \Psi [j - 1] + 1$ where $j = \LF (i)$.  Therefore,
\[\{0\} \cup \{j\ :\ 0 < j \leq n - 1, \Psi [j] \neq \Psi [j - 1] + 1\}\]
also has cardinality $r$ and applying Theorem~\ref{thm:NT21} with $d = 2$ to $\Psi$ splits the runs in the BWT such that no block in $B_F$ overlaps more than 3 blocks in $B_L$, without increasing the number of runs by more than a factor of $3 / 2$.  In fact, the number of runs increases by only 1, from 40 to 41, as shown below with the split block --- corresponding to the first run of 6 copies of {\tt T} in $L$ --- in red:
\begin{eqnarray*}
B_F' & = & \mathtt{1110011011100111111111110001011101101110010100111\textcolor{red}{\bf 100100}10101111000} \\
B_L' & = & \mathtt{\textcolor{red}{\bf 100100}110111011001011111010010011100111000110111111111110001011110}\,.
\end{eqnarray*}

Suppose we apply Theorem~\ref{thm:NT21} with $d = 2$ to $\Psi$ and then store, for $0 \leq b < r$, the index of the block in $B_L$ containing $\LF (i_b)$ and $\LF (i_b)$'s offset in that block, where $i_b$ is the starting position of block $b$ in $B_L$.  Nishimoto and Tabei called this the {\em move table} for $\LF$ (see also~\cite{BGR22,ZBAGL24}) and it takes a total of $O (r \log n)$ bits.  If we know $B_L [j]$ is in block $b$ in $B_L$ with offset $j - i_b$ then, since the block in $B_F$ to which $\LF$ maps block $b$ in $B_L$ now overlaps at most the block containing $B_L [\LF (i_b)]$ and the next 2 blocks in $B_L$, we can find the index of the block in $B_L$ containing $B_L [\LF (j)] = B_L [\LF (i_b)] + j - i_b$ and $B_L [\LF (j)]$'s offset in that block with at most 2 constant-time select queries on $B_L$.  We could use at most 2 constant-time lookups instead if we have the starting positions of the blocks in $B_L$ stored explicitly in another $O (r \log n)$ bits.

\subsection{A faster RLCSA without rank queries}
\label{sec:speedup}

Recall that the $O (\log \log n)$ term in the query-time bound in Corollary~\ref{cor:interpolative} comes only from the use of rank queries on an SD-vector.  Since rank and select queries can be combined to support predecessor queries and select queries on sparse bitvectors can easily be supported in constant time and space polynomial in the number of 1s, rank queries on compact sparse bitvectors inherit lower bounds from predecessor queries~\cite{BF02} --- so they cannot be implemented in constant time.  Therefore, to get rid of that $O (\log \log n)$ term, we must somehow avoid rank queries.

We could replace the rank queries with a move table, but that would result in an $O (r \log n)$ term in our space bound.  Instead, we introduce an uncompressed $2 r$-bit bitvector $B_{FL}$ indicating how the starting positions of the blocks in $F$ and $L$ are interleaved.  Specifically, we scan $B_F$ and $B_L$ simultaneously --- assuming we have already applied Theorem~\ref{thm:NT21} to them so that no block in $F$ overlaps more than 3 blocks in $L$ (so $r$ is a constant factor larger than it was before the application of the theorem) --- and
\begin{itemize}
\item if we see 0s in both bitvectors in position $i$ then we write nothing;
\item if we see a 1 in $B_F$ and a 0 in $B_L$ then we write a 1 (indicating that a block starts in $F$);
\item if we see a 0 in $B_F$ and a 1 in $B_L$ then we write a 0 (indicating that a block starts in $L$);
\item if we see 1s in both bitvectors then we write a 0 and then a 1 (indicating that blocks start in both $L$ and $F$).
\end{itemize}
This way, $B_{FL}.\select_1 (j)$ tells us which at most 3 blocks in $L$ --- those corresponding to the 0 preceding the $j$th copy of 1 in $B_{FL}$ and possibly to the next 2 copies of 0 --- could overlap block $j$ in $F$ (counting from 1).  We can then find the starting positions of those blocks in $L$ using at most 3 select queries on $B_L$.

For our example, taking $B_F$ and $B_L$ to be $B_F'$ and $B_L'$ shown at the end of Section~\ref{sec:NT21},
\begin{eqnarray*}
&   & \mathtt{\textcolor{gray}{0123456789012345678901234567890123456789012345\ 67890123456789012345}} \\
B_F & = & \mathtt{11100110111001111111111100010111011011100\textcolor{red}{\bf 10100}\,11110010010101111000} \\
B_L & = & \mathtt{10010011011101100101111101001001110011\textcolor{blue}{\bf 100011011}1111111110001011110}\,,
\end{eqnarray*}
(with the grey numbers only to show positions) we have
\begin{eqnarray*}
& & \mathtt{\textcolor{gray}{0123456789012345678901234567890123456789}}\\
B_{FL} & = & \mathtt{0111010101010100101110110101010101010110} \ldots \\
& &  \hspace{6ex} \mathtt{\textcolor{gray}{012345678901234\,567890123456789012345678901}} \\
       &   & \hspace{3ex} \ldots \mathtt{100110101\textcolor{blue}{\bf 0}1\textcolor{red}{\bf 1}0\textcolor{blue}{\bf 0}\textcolor{red}{\bf 1}001010101000100011011010100}\,.
\end{eqnarray*}
In position 38 we see 1s in both $B_F$ and $B_L$, so we write 01 in $B_{FL}$ (in positions 49 and 50, respectively); in positions 39 and 40 we see 0s in both in $B_F$ and $B_L$, so we write nothing; in position 41 we see a 1 in $B_F$ and a 0 in $B_L$, so we write a 1 in $B_{FL}$; in position 42 we see a 0 in $B_F$ and a 1 in $B_L$, so we write a 0 in $B_{FL}$; in position 43 we see 1s in both $B_F$ and $B_L$, so we write 01 in $B_{FL}$; in position 44 we see 0s in both $B_F$ and $B_L$, so we write nothing; and in position 45 we see a 0 in $B_F$ and a 1 in $B_L$, so we a 0 in $B_{FL}$ (in position 55).  Admittedly, when $n = 66$ and after applying Theorem~\ref{thm:NT21} $2 r = 82$, it seems foolish to store a $2 r$-bit uncompressed bitvector instead of simply storing $B_L$ uncompressed.  This is due to the small size of our example, however; for massive and highly repetitive datasets, $r$ can easily be hundreds of times smaller than $n$.

Suppose we know the SA interval $\SA [41..43]$ for $a P$ starts at offset 0 in block 28 in $F$ and ends at offset 1 in block 29 in $F$ and we want to find which blocks contain its starting and ending positions in $L$ and the offsets of those positions.  In Section~\ref{sec:review}, we performed 2 rank queries on $B_L$, but now we perform queries $B_{FL}.\select_1 (29) = 51$ and $B_{FL}.\select_1 (30) = 54$ (with arguments 29 and 30 instead of 28 and 29 because we mark with a 1 the starting of the first block in $F$, which we index with 0; the results 51 and 54 are indexed from 0 as well).  Since the 29th and 30th copies of 1 are $B_{FL} [51]$ and $B_{FL} [54]$ (shown in red above), they are preceded by the 23rd and 25th copies of 0, respectively.

Because we applied Theorem~\ref{thm:NT21}, this means the 29th and 30th blocks in $F$ (shown in red in $B_F$ above) overlap the 23rd block in $L$ and possibly the 24th and 25th blocks (shown in blue in $B_L$), and the 25th block and possibly the 26th and 27th blocks (also shown in blue in $B_L$).  Notice that, because we split the 34th block in $F$ but the first block in $L$ for Theorem~\ref{thm:NT21}, the block numbers we find in $F$ are the same as in Section~\ref{sec:review} but the block numbers we find in $L$ will be incremented.  Although in general we need 6 select queries on $B_L$, in this case we can use only 5 --- $B_L.\select_1 (23), \ldots, B_L.\select_1 (27)$ --- to find where these blocks begin in constant time, and determine which contain the starting and ending positions of the SA interval $\SA [41..43]$: the 23rd and the 25th, respectively.

In short, we replace a rank query on SD-bitvector $B_L$ by queries on uncompressed bitvector $B_{FL}$ and constant-time select queries on $B_L$.  This gives us the following theorem:

\begin{theorem}
\label{thm:speedup}
We can store an RLCSA for $T$ in $O \left( \rule{0ex}{2ex} r \log (n / r) + r \log \sigma + \sigma \right)$ bits such that, given character $a$ and the SA interval for $P$, we can find the SA interval for $a P$ in $O (\log r_a)$ time, where $r_a$ is the number of runs of copies of $a$ in the BWT of $T$.
\end{theorem}

\noindent Instead of viewing $B_{FL}$ as replacing slow rank queries while using the overall same space, we can also view it (and $B_F$ and $B_L$) as replacing an $O (r \log n)$-bit move table while using the same overall query time.  Brown, Gagie and Rossi~\cite{BGR22} implemented a similar approach to speeding up $\LF$ computations in an RLFM-index, but only alluded to it briefly in their paper --- the path to {\tt Bitvector} in their Figure 3 --- and gave no analysis nor bounds.  We conjecture that a similar approach can also be applied to reduce the size of fast move tables for $\phi$ and $\phi^{- 1}$~\cite{KMP09}, which return $\SA [i - 1]$ and $\SA [i + 1]$ when given $\SA [i]$.

\section{Two-level indexing}
\label{sec:two-level_indexing}

Corollary~\ref{cor:interpolative} and Theorem~\ref{thm:speedup} suggest that RLCSAs should perform well compared to FM-indexes and RLFM-indexes when the BWT is over a fairly large alphabet and the number of runs of each character is fairly small; Ord{\'o}{\~n}ez, Navarro and Brisaboa~\cite{ONB17} have confirmed this experimentally.  When indexing a highly repetitive text over a small alphabet, we can make RLCSAs more practical by storing a table of $k$-tuples that tells us in which range of $\Psi'$ to search based on which character we are trying to match and which $k - 1$ characters we have just matched.  (This table can be represented with a bitvector to save space.)  The table for our example from Figure~\ref{fig:CSAs} and $k = 2$ is shown below:

\begin{center}
\begin{tabular}{c@{\hspace{5ex}}c@{\hspace{5ex}}c@{\hspace{5ex}}c}
\begin{tabular}{cl}
{\tt {\#}C} & 0      \\
{\tt {\$}C} & 1..2   \\
{\tt {\$}G} & 3      \\
{\tt AA}    & 4      \\
{\tt AC}    & 4..7
\end{tabular}
&
\begin{tabular}{cl}
{\tt AG}    & 8      \\
{\tt AT}    & 9..12  \\
{\tt CA}    & 13..14 \\
{\tt CC}    & 15..16 \\
{\tt CG}    & 17..19
\end{tabular}
&
\begin{tabular}{cl}
{\tt CT}    & 20..24 \\
{\tt GA}    & 25..27 \\
{\tt GC}    & 28..29 \\
{\tt GG}    & 30..31 \\
{\tt GT}    & 32
\end{tabular}
&
\begin{tabular}{cl}
{\tt T{\#}} & 33     \\
{\tt T{\$}} & 33     \\
{\tt TA}    & 34..35 \\
{\tt TG}    & 36..37 \\
{\tt TT}    & 38..39
\end{tabular}
\end{tabular}
\end{center}

\noindent This says that if we want the SA interval for {\tt GCG} and we have just matched the {\tt C}, then we should search in the range $\Psi' [28..29]$.  Notice that the largest range of $\Psi'$ in which we search is now $\Psi' [20..24]$ --- of length 5 --- when we are trying to match a {\tt C} after just matching a {\tt T}; without such a table, the largest range we search is $\Psi' [13..24]$ --- of length 12 --- when trying to match a {\tt C}.

There are interesting cases in which we want to index highly repetitive texts over large alphabets, however.  For example, consider indexing a minimizer digest of a pangenome --- considering minimizers as meta-characters from a large alphabet instead of tuples of characters from a small alphabet~\cite{ARGBL23,AFKLPPP25,EBC21,ZLSAL25} --- or {\em two-level indexing} such a text.  For two-level indexing we build one index for the text and another for a parse of the text; the alphabet of the parse is the dictionary of distinct phrases, which is usually large, but the parse itself is usually much smaller than the text and its BWT is usually still run-length compressible (albeit less than the BWT of the text) when the text is highly repetitive.

Something like two-level indexing was proposed by Deng, Hon, K\"oppl and Sadakane~\cite{DHKS22} but they did not use an index for the text and its absence made their implementation quite slow for all but very long patterns.  Hong, Oliva, K\"oppl, Bannai, Boucher and Gagie~\cite{HOKBBG24} described another approach, which we will review here, but they used standard FM-indexes for the text and the parse instead of RLFM-indexes, so their two-level index was noticeably faster but hundreds of times larger than its competitors.

Consider the 50 similar toy genomes of length 50 each in Figure~\ref{fig:genomes}.  Suppose we parse their concatenation similarly to {\tt rsync}, by inserting a phrase break whenever we see a trigger string --- {\tt ACA}, {\tt ACG}, {\tt CGC}, {\tt CGG}, {\tt GAC}, {\tt GAG}, {\tt GAT}, {\tt GTG}, {\tt GTT}, {\tt TCG} or {\tt TCT} --- or when we reach the {\tt TA\#} at the end of the text.  (Considering $\mathtt{\#} = \mathtt{\$} = 0$, $\mathtt{A} = 1$, $\mathtt{C} = 2$, $\mathtt{G} = 3$ and $\mathtt{T} = 4$ and viewing triples as 3-digit numbers in base 5, the trigger strings are the triples in the concatenation whose values are congruent to 0 modulo 6.)  If we replace each phrase in the parse by its lexicographic rank in the dictionary of distinct phrases, counting from 1, and terminate the sequence with a 0, we get the 562-number sequence shown in Figure~\ref{fig:parse}.  With a larger example, of course, we obtain longer phrases on average and better compression from the parsing.

\begin{figure}[t!]
\resizebox{\textwidth}{!}
{\tt \begin{tabular}{cc}
\begin{tabular}{c}
CTTCCGCGGTGATAAAGGGGGCGGTAATGTCGCGAAACAGTCTTTTCTA\$ \\
CTTACGCGGTGATACAGGGGGCCGTAATTTCGCGGAACAGTCTTTTCTA\$ \\
CTTACGCGACGATCCAGGGGGCGGTAATTTCGCGGAACAGTCTTTTCTA\$ \\
CTTATGCGATGATCCTGGGGGCGGTAATTTCGCGGAACAGTCTTTTCTA\$ \\
CTTACGCGGTGATCCAGGGGGCGGTAATTTCGCGGAACACTCTTCTCTA\$ \\
CTTACGCGATGATCCAGTGGGCGGTCTTTTCGCGGAACAGTCTTTTCGA\$ \\
CTTACGCGGTGATCCAGGGGGCGGTAATTTCGCGCAACAGTCTTTTCTA\$ \\
CTTACGCGGTGATCCAGGGGGCGGTAATTTCTCGGAACAGTCTTTTCTA\$ \\
CTTATGCGGTGATCCACGGGGCGGAAGTATCGCGGAACAGTCTTTTTTA\$ \\
CTTACGCGATGATCCAGGGGGCGGTAACTTCGCGGAACAGTCTTTTCTA\$ \\
CTTACGCGACGATCCAGGGGGCAGTAATTTCGCGGAACAGTCTTTTCTA\$ \\
CATACGCGGGGATCCAAGGGGCGGTAATTTCGCGGAACAGTCTTTGACA\$ \\
CTTTCACGGTGATCCAGGGGTGGGTAATTTCGCGGAACAGTCTTTTCTA\$ \\
CTTACGCGGTGATCCAGGGGGCGGTAATTTCGCGGAACAATCTTTTCTA\$ \\
CTTACGCGATGATCCAGGGGGCGGTAATTTCGCGGAACAGTCTTTTCTA\$ \\
CTTACGCGGTGATCCAGGGGGCGCTAATTTCGCGAAACAGTCTTTTCTA\$ \\
CTTACGTGGTGATCCAGGGGGCGGTAATTTCGAGGAACAGTCTTTAATA\$ \\
CTTACGCGGTGATCCAGGGCGCGGTAATTTCGCGGAACAATCTTTTCTA\$ \\
CTTACGCGATGATCCAGGGGGCGGTAATTTCGCGGAACAGTCTAATCTA\$ \\
CTTACGCGGTGATCCAGGGGGCGGTAATTTCGCGGAACAGTCTTTTCTA\$ \\
CTTACGCGGTGATCCAGGGGGCGGTAATTTCGCGGAACAGTCTTTTCTA\$ \\
CTTACGCGGTGATCCAGGGGGCGGTAATTTCGCGGAACAATCTTTTCTA\$ \\
CTTACGCGATGATCCAGGGGGCGGTAATTGCGCGGAACAGTCTTTTCTA\$ \\
CTTACGCGGTGATCCAGGGGGCGGTAATTTCGGGGAACAGTCTTTTCTA\$ \\
CTTACGCGATCTTCCAGGGGGCCGAAATTTCGCGTAACAGTCTTTTCTA\$
\end{tabular}
&
\begin{tabular}{c}
CTTACGCGGTGATTCAGGGGGCGGTAATTTCGCGGATCAGTCTTTTCTA\$ \\
CTTACGCGGTTATCCAGGGGGTGGTACTTTCGGTGAACAGTCTGTTCTA\$ \\
CTTACGCGGTGATCCAGGGGGCAGTAATTTCGCGGAACAGTCTTTTCTA\$ \\
CTTACGCGATGATCCAGGGGGCGGTAATTTCGCTGAACAGTCTTTTCTA\$ \\
CTTACGCGATGATCCATTGGGCGGTAATTCCGCGGAACAGTCTTTTCTA\$ \\
CTTACGCGATGATCCATGGGGCGGTAATTTCGCGGAACAGTCTTTTCTA\$ \\
CTTACGCGATGATCCAGGGGGCTGTATTTTCGCGGAACAGTCTTTTCTA\$ \\
CTTACGCGCTGATCCAGGGGGCGGTAATTTCGCGGAACAGTCTTTTCTA\$ \\
CTTACGAGATGAGCTAGGGGGCGGTAATTTCGCGGAACAGTCTTTTCTA\$ \\
CTTACGCGATGCTCCAGGGGGCGGTGATTTCGCGGAACAGTCTTTTCTA\$ \\
CTTTCGCGATGATCCAGGGGGCGGTCATTTCGCGGAACAGTCTTTTCTA\$ \\
CTTACGCGGTGATCCAGGGGGCGGTAATTTCGCGGCACAGTCCTTTCTA\$ \\
CTTACGCGGTGATCCAGGGGGCGGTAATTTCGCGGAACAGTCTTTTCTA\$ \\
CTTACGCGGTGATCCAGGGGGCGGTAATTTCGCGGAACAGTCTTTTCTA\$ \\
CTTACTCGGTGATCCAGGGGGCGGTAATTTCGCGGAACAGTCTTTTCTT\$ \\
CTTACGCGATGATCCAGAGGGCGGTAATTTCGCGGAACAGTCTTTTCTA\$ \\
CTTACGCGGTGATCCAGGGGGCGGTAATTCCGCGGAACAGTCTTTTCTA\$ \\
CTTACGCGGGGATCCAGGGGGCGGTAATTTCGCGGAACAATCTTTTCTA\$ \\
CTTACGCGGTGATCCAGGGGTCGGTAATTTCGCGGAACAGTCTTTTCTA\$ \\
CATACGCGGTGATCCAGGGGGCGGTAATTTCGCGGAACAGTCTTTTCTA\$ \\
CTTACGCGGTGATCCAGGGGGCGGTAATTTCGCGGAACAGTCTTTTCTA\$ \\
CCTACGCGATGATGCAGGGGGCGGTAATTTCGCGGAACAGTCTTTCCTA\$ \\
CTTACGCGATGTTCCAGGGGGCGGTAATTTCGCGGAATAGTTTTTTCTA\$ \\
CTTACGCGGTGATCCAGCGGGCGGTAATTTCGCGGAATAGTCTTTTCTA\$ \\
CTTACGCGGTAATCCAGGGGGCGGTAATGTCGCGGAACAGTCTTTTCTA\#
\end{tabular}
\end{tabular}}
\caption{A set of 50 similar toy genomes of length 50 each, with the first 49 separated by copies of {\tt \$} and the last one terminated by {\tt \#}.}
\label{fig:genomes}
\end{figure}

\begin{figure}[t!]
\resizebox{\textwidth}{!}
{\begin{tabular}{rrrrrrrrrrrrrrrrrrrr}
44, & 55, & 79, & 19, & 11, & 70, & 22, & 46, & 64, & 88, &  6, & 22, & 55, & 79, & 19, & 17, & 59, & 22, & 55, & 12, \\
64, & 88, &  6, & 22, & 48, & 45, & 19, & 32, & 73, & 22, & 55, & 12, & 64, & 88, &  8, & 50, & 39, & 73, & 22, & 55, \\
12, & 64, & 88, &  6, & 22, & 55, & 79, & 19, & 32, & 73, & 22, & 55, & 12, & 43, & 78, & 41, &  6, & 622, & 50, & 50, \\
36, & 58, & 78, & 87, & 22, & 55, & 12, & 64, & 87, &  6, & 22, & 55, & 79, & 19, & 32, & 73, & 22, & 51, & 12, & 64, \\
88, &  6, & 22, & 55, & 79, & 19, & 32, & 74, & 40, & 45, & 12, & 64, & 88, &  9, & 79, & 19, & 26, & 45, & 58, & 13, \\
22, & 55, & 12, & 64, & 90, & 22, & 50, & 50, & 32, & 68, & 22, & 55, & 12, & 64, & 88, &  6, & 22, & 48, & 45, & 19, \\
30, & 22, & 55, & 12, & 64, & 88, &  4, & 22, & 55, & 57, & 25, & 73, & 22, & 55, & 12, & 64, & 86, &  2, &  1, & 45, \\
79, & 19, & 35, & 60, & 22, & 55, & 12, & 64, & 88, &  6, & 22, & 55, & 79, & 19, & 32, & 73, & 22, & 55, & 12, & 21, \\
88, &  6, & 22, & 50, & 50, & 32, & 73, & 22, & 55, & 12, & 64, & 88, &  6, & 22, & 55, & 79, & 19, & 31, & 73, & 22, \\
46, & 64, & 88, &  6, & 79, & 65, & 19, & 32, & 73, & 18, & 47, & 64, & 83, & 22, & 55, & 79, & 19, & 29, & 55, & 73, \\
22, & 55, & 12, & 21, & 88, &  6, & 22, & 50, & 50, & 32, & 73, & 22, & 55, & 12, & 64, & 16, &  6, & 22, & 55, & 79, \\
19, & 32, & 73, & 22, & 55, & 12, & 64, & 88, &  6, & 22, & 55, & 79, & 19, & 32, & 73, & 22, & 55, & 12, & 64, & 88, \\
 6, & 22, & 55, & 79, & 19, & 32, & 73, & 22, & 55, & 12, & 21, & 88, &  6, & 22, & 50, & 50, & 32, & 72, & 55, & 12, \\
64, & 88, &  6, & 22, & 55, & 79, & 19, & 32, & 73, & 45, & 56, & 64, & 88, &  6, & 22, & 50, & 41, & 77, & 22, & 61, \\
64, & 88, &  6, & 22, & 55, & 79, & 19, & 75, & 73, & 22, & 55, & 19, & 24, & 88, &  6, & 22, & 55, & 82, & 20, & 62, \\
45, & 79, & 12, & 64, & 66, & 41, &  6, & 22, & 55, & 79, & 19, & 30, & 22, & 55, & 12, & 64, & 88, &  6, & 22, & 50, \\
50, & 32, & 73, & 22, & 80, & 64, & 88, &  6, & 22, & 50, & 50, & 38, & 71, & 55, & 12, & 64, & 88, &  6, & 22, & 50, \\
50, & 37, & 73, & 22, & 55, & 12, & 64, & 88, &  6, & 22, & 50, & 50, & 33, & 22, & 55, & 12, & 64, & 88, &  6, & 22, \\
51, & 81, & 32, & 73, & 22, & 55, & 12, & 64, & 88, &  6, & 18, & 19, & 49, & 42, & 73, & 22, & 55, & 12, & 64, & 88, \\
 6, & 22, & 50, & 54, & 79, & 19, & 85, & 22, & 55, & 12, & 64, & 88, & 10, & 22, & 50, & 50, & 32, & 76, & 22, & 55, \\
12, & 64, & 88, &  6, & 22, & 55, & 79, & 19, & 32, & 73, & 22, & 55, & 23, & 63, &  6, & 22, & 55, & 79, & 19, & 32, \\
73, & 22, & 55, & 12, & 64, & 88, &  6, & 22, & 55, & 79, & 19, & 32, & 73, & 22, & 55, & 12, & 64, & 88, &  7, & 45, \\
79, & 19, & 32, & 73, & 22, & 55, & 12, & 64, & 88, & 67, & 22, & 50, & 50, & 27, & 58, & 73, & 22, & 55, & 12, & 64, \\
88, &  6, & 22, & 55, & 79, & 19, & 32, & 71, & 55, & 12, & 64, & 88, &  6, & 22, & 55, & 57, & 32, & 73, & 22, & 55, \\
12, & 21, & 88, &  6, & 22, & 55, & 79, & 19, & 34, & 45, & 73, & 22, & 55, & 12, & 64, & 88, &  4, & 22, & 55, & 79, \\
19, & 32, & 73, & 22, & 55, & 12, & 64, & 88, &  6, & 22, & 55, & 79, & 19, & 32, & 73, & 22, & 55, & 12, & 64, & 88, \\
 5, & 22, & 50, & 50, & 52, & 73, & 22, & 55, & 12, & 64, & 84, & 22, & 50, & 66, & 32, & 73, & 22, & 55, & 15, & 89, \\
 6, & 22, & 55, & 79, & 19, & 28, & 53, & 73, & 22, & 55, & 14, & 88, &  6, & 22, & 55, & 69, & 70, & 22, & 55, & 12, \\
64, & 88, &  3, &  0
\end{tabular}}
\caption{The 563-number sequence (20 numbers per line) over the alphabet $\{0, \ldots, 90\}$ we get from the concatenation of the toy genomes in Figure~\ref{fig:genomes} by parsing, replacing each phrase by its rank in the dictionary (counting from 1) and appending a 0.}
\label{fig:parse}
\end{figure}

Run-length compression naturally works better on the BWT of the concatenation of the genomes than on the BWT of the parse, as shown in Figures~\ref{fig:genomes_RLBWT} and~\ref{fig:parse_RLBWT}.  Again, with a larger example we would achieve better compression, also from the run-length compressed BWT (RLBWT) of the parse.  Even this small example, however, gives some intuition how the dictionary of distinct phrases in the parse is usually large, but the parse is usually much smaller than the text and its BWT is usually still run-length compressible.  In this case there are 90 distinct phrases in the dictionary, the parse is less than a quarter as long as the text, and the average run length in its RLBWT is slightly more than 2.  An FM-index based on the RLBWT of the parse would generally use at least about $\lceil \lg 90 \rceil = 7$ rank queries on bitvectors for each backward step.  The most common value in the runs, 19, occurs in only 16 runs, so we should spend at most about $\lg 16 = 4$ steps in each binary search.

\begin{figure}[t!]
\resizebox{\textwidth}{!}
{\begin{tabular}{cccccccccccccccccccc}
$\mathtt{A}^{8}$ & $\mathtt{T}^{1}$ & $\mathtt{A}^{41}$ & $\mathtt{T}^{28}$ & $\mathtt{G}^{1}$ & $\mathtt{T}^{18}$ & $\mathtt{C}^{1}$ & $\mathtt{T}^{1}$ & $\mathtt{G}^{2}$ & $\mathtt{T}^{1}$ & $\mathtt{G}^{27}$ & $\mathtt{A}^{1}$ & $\mathtt{G}^{10}$ & $\mathtt{C}^{1}$ & $\mathtt{T}^{1}$ & $\mathtt{A}^{1}$ & $\mathtt{G}^{6}$ & $\mathtt{T}^{1}$ & $\mathtt{C}^{1}$ \\
$\mathtt{A}^{1}$ & $\mathtt{G}^{1}$ & $\mathtt{T}^{1}$ & $\mathtt{G}^{2}$ & $\mathtt{T}^{2}$ & $\mathtt{C}^{4}$ & $\mathtt{T}^{39}$ & $\mathtt{A}^{1}$ & $\mathtt{T}^{3}$ & $\mathtt{G}^{1}$ & $\mathtt{A}^{5}$ & $\mathtt{T}^{1}$ & $\mathtt{C}^{1}$ & $\mathtt{A}^{41}$ & $\mathtt{T}^{1}$ & $\mathtt{G}^{2}$ & $\mathtt{T}^{42}$ & $\mathtt{C}^{2}$ & $\mathtt{T}^{2}$ & $\mathtt{C}^{1}$ \\
$\mathtt{A}^{1}$ & $\mathtt{T}^{1}$ & $\mathtt{C}^{1}$ & $\mathtt{G}^{1}$ & $\mathtt{C}^{1}$ & $\mathtt{G}^{2}$ & $\mathtt{C}^{1}$ & $\mathtt{G}^{1}$ & $\mathtt{A}^{1}$ & $\mathtt{C}^{6}$ & $\mathtt{A}^{1}$ & $\mathtt{C}^{13}$ & $\mathtt{T}^{1}$ & $\mathtt{C}^{22}$ & $\mathtt{A}^{1}$ & $\mathtt{C}^{22}$ & $\mathtt{T}^{1}$ & $\mathtt{C}^{22}$ & $\mathtt{T}^{1}$ & $\mathtt{A}^{1}$ \\
$\mathtt{G}^{2}$ & $\mathtt{C}^{2}$ & $\mathtt{A}^{2}$ & $\mathtt{G}^{10}$ & $\mathtt{A}^{1}$ & $\mathtt{G}^{25}$ & $\mathtt{T}^{1}$ & $\mathtt{G}^{6}$ & $\mathtt{T}^{1}$ & $\mathtt{A}^{1}$ & $\mathtt{G}^{1}$ & $\mathtt{A}^{4}$ & $\mathtt{G}^{15}$ & $\mathtt{T}^{2}$ & $\mathtt{G}^{1}$ & $\mathtt{C}^{1}$ & $\mathtt{A}^{2}$ & $\mathtt{G}^{2}$ & $\mathtt{A}^{3}$ & $\mathtt{C}^{1}$ \\
$\mathtt{A}^{27}$ & $\mathtt{C}^{1}$ & $\mathtt{A}^{2}$ & $\mathtt{G}^{1}$ & $\mathtt{A}^{9}$ & $\mathtt{T}^{1}$ & $\mathtt{A}^{1}$ & $\mathtt{G}^{1}$ & $\mathtt{C}^{1}$ & $\mathtt{A}^{4}$ & $\mathtt{G}^{1}$ & $\mathtt{C}^{1}$ & $\mathtt{T}^{1}$ & $\mathtt{A}^{1}$ & $\mathtt{C}^{6}$ & $\mathtt{A}^{1}$ & $\mathtt{C}^{12}$ & $\mathtt{G}^{1}$ & $\mathtt{C}^{11}$ & $\mathtt{T}^{1}$ \\
$\mathtt{C}^{10}$ & $\mathtt{G}^{2}$ & $\mathtt{A}^{35}$ & $\mathtt{T}^{1}$ & $\mathtt{A}^{7}$ & $\mathtt{C}^{1}$ & $\mathtt{\$}^{2}$ & $\mathtt{C}^{2}$ & $\mathtt{T}^{45}$ & $\mathtt{G}^{1}$ & $\mathtt{T}^{3}$ & $\mathtt{G}^{1}$ & $\mathtt{T}^{1}$ & $\mathtt{\$}^{1}$ & $\mathtt{T}^{3}$ & $\mathtt{G}^{2}$ & $\mathtt{C}^{1}$ & $\mathtt{G}^{2}$ & $\mathtt{A}^{1}$ & $\mathtt{T}^{1}$ \\
$\mathtt{A}^{2}$ & $\mathtt{G}^{17}$ & $\mathtt{T}^{2}$ & $\mathtt{A}^{9}$ & $\mathtt{T}^{1}$ & $\mathtt{A}^{7}$ & $\mathtt{T}^{1}$ & $\mathtt{A}^{1}$ & $\mathtt{T}^{18}$ & $\mathtt{C}^{1}$ & $\mathtt{T}^{2}$ & $\mathtt{C}^{1}$ & $\mathtt{T}^{9}$ & $\mathtt{G}^{1}$ & $\mathtt{T}^{9}$ & $\mathtt{A}^{3}$ & $\mathtt{G}^{1}$ & $\mathtt{C}^{1}$ & $\mathtt{A}^{22}$ & $\mathtt{T}^{1}$ \\
$\mathtt{G}^{1}$ & $\mathtt{T}^{1}$ & $\mathtt{G}^{35}$ & $\mathtt{T}^{1}$ & $\mathtt{G}^{9}$ & $\mathtt{T}^{1}$ & $\mathtt{G}^{2}$ & $\mathtt{A}^{1}$ & $\mathtt{G}^{17}$ & $\mathtt{T}^{1}$ & $\mathtt{G}^{23}$ & $\mathtt{T}^{1}$ & $\mathtt{G}^{18}$ & $\mathtt{T}^{1}$ & $\mathtt{G}^{4}$ & $\mathtt{A}^{1}$ & $\mathtt{G}^{4}$ & $\mathtt{C}^{1}$ & $\mathtt{A}^{1}$ & $\mathtt{T}^{17}$ \\
$\mathtt{C}^{1}$ & $\mathtt{T}^{28}$ & $\mathtt{G}^{1}$ & $\mathtt{C}^{1}$ & $\mathtt{G}^{2}$ & $\mathtt{T}^{1}$ & $\mathtt{A}^{1}$ & $\mathtt{T}^{1}$ & $\mathtt{A}^{1}$ & $\mathtt{G}^{2}$ & $\mathtt{C}^{1}$ & $\mathtt{G}^{1}$ & $\mathtt{T}^{2}$ & $\mathtt{\$}^{44}$ & $\mathtt{T}^{1}$ & $\mathtt{\#}^{1}$ & $\mathtt{A}^{1}$ & $\mathtt{T}^{2}$ & $\mathtt{\$}^{1}$ & $\mathtt{T}^{1}$ \\
$\mathtt{\$}^{1}$ & $\mathtt{A}^{1}$ & $\mathtt{C}^{1}$ & $\mathtt{T}^{44}$ & $\mathtt{C}^{4}$ & $\mathtt{G}^{6}$ & $\mathtt{T}^{1}$ & $\mathtt{G}^{15}$ & $\mathtt{T}^{1}$ & $\mathtt{G}^{22}$ & $\mathtt{T}^{1}$ & $\mathtt{C}^{3}$ & $\mathtt{T}^{1}$ & $\mathtt{C}^{1}$ & $\mathtt{A}^{1}$ & $\mathtt{T}^{2}$ & $\mathtt{G}^{2}$ & $\mathtt{T}^{4}$ & $\mathtt{C}^{1}$ & $\mathtt{T}^{9}$ \\
$\mathtt{G}^{1}$ & $\mathtt{T}^{10}$ & $\mathtt{C}^{1}$ & $\mathtt{T}^{13}$ & $\mathtt{C}^{1}$ & $\mathtt{A}^{1}$ & $\mathtt{C}^{13}$ & $\mathtt{T}^{1}$ & $\mathtt{C}^{2}$ & $\mathtt{T}^{2}$ & $\mathtt{C}^{1}$ & $\mathtt{G}^{1}$ & $\mathtt{T}^{1}$ & $\mathtt{G}^{4}$ & $\mathtt{C}^{16}$ & $\mathtt{T}^{1}$ & $\mathtt{C}^{4}$ & $\mathtt{T}^{1}$ & $\mathtt{G}^{2}$ & $\mathtt{C}^{38}$ \\
$\mathtt{G}^{1}$ & $\mathtt{C}^{4}$ & $\mathtt{A}^{1}$ & $\mathtt{C}^{2}$ & $\mathtt{G}^{1}$ & $\mathtt{C}^{1}$ & $\mathtt{G}^{8}$ & $\mathtt{C}^{1}$ & $\mathtt{G}^{29}$ & $\mathtt{C}^{2}$ & $\mathtt{T}^{1}$ & $\mathtt{C}^{20}$ & $\mathtt{G}^{1}$ & $\mathtt{C}^{3}$ & $\mathtt{A}^{1}$ & $\mathtt{T}^{1}$ & $\mathtt{C}^{2}$ & $\mathtt{G}^{1}$ & $\mathtt{C}^{6}$ & $\mathtt{A}^{1}$ \\
$\mathtt{C}^{10}$ & $\mathtt{G}^{1}$ & $\mathtt{C}^{26}$ & $\mathtt{G}^{2}$ & $\mathtt{C}^{1}$ & $\mathtt{G}^{54}$ & $\mathtt{A}^{1}$ & $\mathtt{G}^{5}$ & $\mathtt{T}^{1}$ & $\mathtt{G}^{16}$ & $\mathtt{A}^{1}$ & $\mathtt{G}^{8}$ & $\mathtt{C}^{1}$ & $\mathtt{G}^{7}$ & $\mathtt{T}^{1}$ & $\mathtt{G}^{2}$ & $\mathtt{C}^{3}$ & $\mathtt{G}^{5}$ & $\mathtt{C}^{1}$ & $\mathtt{G}^{13}$ \\
$\mathtt{A}^{1}$ & $\mathtt{G}^{2}$ & $\mathtt{T}^{1}$ & $\mathtt{G}^{19}$ & $\mathtt{A}^{23}$ & $\mathtt{T}^{1}$ & $\mathtt{A}^{18}$ & $\mathtt{G}^{1}$ & $\mathtt{T}^{1}$ & $\mathtt{G}^{3}$ & $\mathtt{C}^{25}$ & $\mathtt{G}^{1}$ & $\mathtt{C}^{14}$ & $\mathtt{T}^{1}$ & $\mathtt{C}^{1}$ & $\mathtt{G}^{1}$ & $\mathtt{C}^{10}$ & $\mathtt{T}^{1}$ & $\mathtt{C}^{18}$ & $\mathtt{G}^{2}$ \\
$\mathtt{C}^{2}$ & $\mathtt{G}^{17}$ & $\mathtt{A}^{1}$ & $\mathtt{G}^{3}$ & $\mathtt{C}^{1}$ & $\mathtt{A}^{1}$ & $\mathtt{G}^{21}$ & $\mathtt{A}^{1}$ & $\mathtt{T}^{1}$ & $\mathtt{G}^{1}$ & $\mathtt{A}^{1}$ & $\mathtt{T}^{2}$ & $\mathtt{G}^{1}$ & $\mathtt{A}^{6}$ & $\mathtt{G}^{1}$ & $\mathtt{A}^{37}$ & $\mathtt{G}^{28}$ & $\mathtt{A}^{1}$ & $\mathtt{G}^{2}$ & $\mathtt{C}^{1}$ \\
$\mathtt{G}^{1}$ & $\mathtt{T}^{2}$ & $\mathtt{A}^{1}$ & $\mathtt{T}^{1}$ & $\mathtt{C}^{8}$ & $\mathtt{T}^{1}$ & $\mathtt{C}^{15}$ & $\mathtt{A}^{1}$ & $\mathtt{C}^{22}$ & $\mathtt{A}^{1}$ & $\mathtt{G}^{2}$ & $\mathtt{T}^{1}$ & $\mathtt{G}^{1}$ & $\mathtt{C}^{1}$ & $\mathtt{G}^{6}$ & $\mathtt{C}^{1}$ & $\mathtt{G}^{35}$ & $\mathtt{A}^{1}$ & $\mathtt{T}^{14}$ & $\mathtt{C}^{1}$ \\
$\mathtt{T}^{3}$ & $\mathtt{A}^{1}$ & $\mathtt{T}^{14}$ & $\mathtt{A}^{1}$ & $\mathtt{T}^{11}$ & $\mathtt{G}^{1}$ & $\mathtt{C}^{1}$ & $\mathtt{A}^{2}$ & $\mathtt{T}^{1}$ & $\mathtt{G}^{1}$ & $\mathtt{T}^{2}$ & $\mathtt{G}^{1}$ & $\mathtt{T}^{2}$ & $\mathtt{A}^{1}$ & $\mathtt{G}^{1}$ & $\mathtt{A}^{7}$ & $\mathtt{T}^{1}$ & $\mathtt{A}^{22}$ & $\mathtt{T}^{1}$ & $\mathtt{A}^{5}$ \\
$\mathtt{C}^{1}$ & $\mathtt{A}^{7}$ & $\mathtt{T}^{4}$ & $\mathtt{A}^{1}$ & $\mathtt{G}^{1}$ & $\mathtt{T}^{2}$ & $\mathtt{G}^{1}$ & $\mathtt{T}^{12}$ & $\mathtt{G}^{1}$ & $\mathtt{T}^{23}$ & $\mathtt{A}^{1}$ & $\mathtt{T}^{6}$ & $\mathtt{C}^{1}$ & $\mathtt{T}^{1}$ & $\mathtt{G}^{1}$ & $\mathtt{T}^{1}$ & $\mathtt{C}^{1}$ & $\mathtt{T}^{13}$ & $\mathtt{C}^{1}$ & $\mathtt{T}^{16}$ \\
$\mathtt{A}^{1}$ & $\mathtt{T}^{13}$ & $\mathtt{G}^{1}$ & $\mathtt{T}^{2}$ & $\mathtt{G}^{1}$ & $\mathtt{T}^{1}$ & $\mathtt{A}^{1}$ & $\mathtt{C}^{1}$ & $\mathtt{G}^{13}$ & $\mathtt{A}^{3}$ & $\mathtt{G}^{20}$ & $\mathtt{A}^{1}$ & $\mathtt{G}^{10}$ & $\mathtt{C}^{1}$ & $\mathtt{T}^{1}$ & $\mathtt{A}^{1}$ & $\mathtt{G}^{3}$ & $\mathtt{A}^{1}$ & $\mathtt{G}^{4}$ & $\mathtt{A}^{1}$ \\
$\mathtt{G}^{1}$ & $\mathtt{A}^{1}$ & $\mathtt{G}^{5}$ & $\mathtt{A}^{1}$ & $\mathtt{G}^{1}$ & $\mathtt{C}^{1}$ & $\mathtt{A}^{1}$ & $\mathtt{G}^{7}$ & $\mathtt{A}^{1}$ & $\mathtt{G}^{2}$ & $\mathtt{A}^{2}$ & $\mathtt{G}^{2}$ & $\mathtt{A}^{5}$ & $\mathtt{G}^{2}$ & $\mathtt{A}^{2}$ & $\mathtt{T}^{1}$ & $\mathtt{A}^{2}$ & $\mathtt{T}^{1}$ & $\mathtt{G}^{1}$ & $\mathtt{A}^{1}$ \\
$\mathtt{C}^{1}$ & $\mathtt{G}^{3}$ & $\mathtt{C}^{1}$ & $\mathtt{A}^{3}$ & $\mathtt{C}^{2}$ & $\mathtt{T}^{2}$ & $\mathtt{C}^{42}$ & $\mathtt{G}^{1}$ & $\mathtt{C}^{2}$ & $\mathtt{T}^{1}$ & $\mathtt{A}^{1}$ & $\mathtt{C}^{1}$ & $\mathtt{G}^{1}$ & $\mathtt{A}^{2}$ & $\mathtt{C}^{1}$ & $\mathtt{T}^{19}$ & $\mathtt{C}^{1}$ & $\mathtt{T}^{47}$ & $\mathtt{G}^{1}$ & $\mathtt{T}^{21}$ \\
$\mathtt{C}^{1}$ & $\mathtt{T}^{2}$ & $\mathtt{A}^{2}$ & $\mathtt{T}^{1}$ & $\mathtt{C}^{3}$ & $\mathtt{T}^{1}$ & $\mathtt{A}^{2}$ & $\mathtt{C}^{1}$ & $\mathtt{A}^{9}$ & $\mathtt{T}^{1}$ & $\mathtt{A}^{8}$ & $\mathtt{T}^{1}$ & $\mathtt{A}^{20}$ & $\mathtt{C}^{1}$ & $\mathtt{T}^{28}$ & $\mathtt{C}^{1}$ & $\mathtt{T}^{12}$ & $\mathtt{A}^{1}$ & $\mathtt{T}^{1}$ & $\mathtt{C}^{1}$ \\
$\mathtt{T}^{1}$ & $\mathtt{C}^{2}$ & $\mathtt{A}^{1}$ & $\mathtt{C}^{20}$ & $\mathtt{T}^{1}$ & $\mathtt{C}^{20}$ & $\mathtt{T}^{2}$ & $\mathtt{C}^{1}$ & $\mathtt{G}^{1}$ 
\end{tabular}}
\caption{The RLBWT of the concatenation of the toy genomes shown in Figure~\ref{fig:genomes}, consisting of 449 runs (20 runs per line).}
\label{fig:genomes_RLBWT}
\end{figure}

\begin{figure}[b!]
\resizebox{\textwidth}{!}
{\begin{tabular}{rrrrrrrrrrrrrrr}
$3$, & $2$, & $86$, & $88^{12}$, & $41$, & $88^{6}$, & $89$, & $41$, & $88$, & $87$, & $88^{4}$, & $16$, & $63$, & $88^{11}$, & $19$, \\
$55^{5}$, & $79$, & $55^{25}$, & $51$, & $55$, & $45$, & $55^{2}$, & $58$, & $55$, & $64$, & $19$, & $6$, & $73$, & $79$, & $55$, \\
$79^{2}$, & $45$, & $79^{2}$, & $65$, & $79^{9}$, & $45$, & $79^{5}$, & $18$, & $79$, & $82$, & $12^{3}$, & $70$, & $73$, & $6^{2}$, & $67$, \\
$90$, & $6^{3}$, & $10$, & $6^{3}$, & $5$, & $6$, & $84$, & $73$, & $6$, & $73^{7}$, & $87$, & $70$, & $30$, & $73^{2}$, & $68$, \\
$59$, & $30$, & $73$, & $33$, & $73^{2}$, & $60$, & $73^{3}$, & $76$, & $73$, & $85$, & $73$, & $13$, & $73^{3}$, & $4$, & $6^{3}$, \\
$83$, & $6^{8}$, & $4$, & $6^{8}$, & $77$, & $73$, & $55$, & $19$, & $57$, & $19$, & $50$, & $19^{4}$, & $50$, & $19$, & $50$, \\
$19^{3}$, & $57$, & $19$, & $50$, & $19$, & $81$, & $50$, & $19^{6}$, & $66$, & $19$, & $50$, & $19$, & $50$, & $19$, & $50^{3}$, \\
$74$, & $78$, & $66$, & $50$, & $49$, & $12$, & $0$, & $40$, & $48$, & $73$, & $26$, & $34$, & $62$, & $7$, & $1$, \\
$22$, & $18$, & $22$, & $19$, & $50^{10}$, & $8$, & $22^{12}$, & $50$, & $22^{3}$, & $50$, & $28$, & $50$, & $22^{17}$, & $71$, & $22$, \\
$71$, & $22^{8}$, & $72$, & $22^{11}$, & $29$, & $44$, & $22^{21}$, & $45$, & $55$, & $45$, & $27$, & $36$, & $17$, & $35$, & $22$, \\
$20$, & $23$, & $12$, & $47$, & $12^{9}$, & $56$, & $12^{2}$, & $80$, & $12^{2}$, & $46$, & $12^{10}$, & $61$, & $46$, & $12^{5}$, & $79$, \\
$50$, & $64$, & $88$, & $32$, & $55$, & $11$, & $69$, & $38$, & $32^{2}$, & $31$, & $32$, & $55$, & $32^{3}$, & $52$, & $25$, \\
$45$, & $32$, & $37$, & $42$, & $32$, & $58$, & $32$, & $39$, & $32^{5}$, & $53$, & $32$, & $75$, & $32^{3}$, & $19$, & $32$, \\
$41$, & $43$, & $58$, & $45$, & $55$, & $9$, & $55^{14}$, & $45$, & $55^{3}$, & $45$, & $55$, & $54$, & $6$, & $22$, & $51$, \\
$55$, & $64$, & $19$, & $64$, & $78$, & $64^{7}$, & $21^{2}$, & $64^{6}$, & $14$, & $64^{11}$, & $21$, & $64$, & $24$, & $64^{5}$, & $15$, \\
$64$
\end{tabular}}
\caption{The RLBWT of the sequence shown in Figure~\ref{fig:parse_RLBWT}, consisting of 226 runs (15 runs per line).}
\label{fig:parse_RLBWT}
\end{figure}

To search for a pattern, we start by backward stepping in the index for the text until we reach the left end of the rightmost trigger string in the pattern.  We keep count of how often each trigger string occurs in the text and an $n$-bit sparse bitvector with 1s marking the lexicographic ranks of the lexicographically least suffixes starting with each trigger string.  This way, when we reach the left end of the rightmost trigger string, with a rank query on that bitvector we can compute the lexicographic ranks of the suffixes starting with the suffix of the pattern we have processed so far among all the suffixes starting with trigger strings, and map from the index of the text into the index for the parse.  The width of the BWT interval stays the same and may span several lexicographically consecutive phrases in the dictionary --- all those starting with the suffix of the pattern we have processed so far --- but it is possible to start a backward search in the index for the parse with a lexicographic range of phrases rather than with a single phrase.

When we reach the left end of the leftmost trigger string in the pattern, we can use the same bitvector to map back into the index for the text and match the remaining prefix of the pattern with that.  While matching the pattern phrase by phrase against the index for the parse, we can either compare against phrases in the stored dictionary or just use Karp-Rabin hashes (allowing some probability of false-positive matches).  We still have to parse the pattern, but that requires a single sequential pass, while FM-indexes in particular are known for poor memory locality.  They key idea is that, ideally, we match most of the pattern phrase by phrase instead of character by character, reducing the number of cache misses.

We plan to reimplement two-level indexes for collections of similar genomes with RLFM-indexes for the collections themselves and CSAs, standard RLCSAs and our sped-up RLCSAs for the parses from Theorem~\ref{thm:speedup} of those collections, and compare them experimentally.  We also plan to try indexing minimizer digests with CSAs and RLCSAs.

\section{Boyer-Moore-Li with two-level indexing}
\label{sec:MEM-finding}

Olbrich, B\"uchler and Ohlebusch~\cite{OBO25} recently showed how working with {\tt rsync}-like parses of genomes instead of the genomes themselves can speed up multiple alignment.  More specfically, they find and use as anchors finding maximal substrings (call multi-MUMs) of the parses that occur exactly once in each parse.  In this section we speculate about how two-level indexing may similarly speed up searches for long maximal exact matches (MEMs).  A MEM of a pattern $P [0..m - 1]$ with respect to a text $T$ is a substring $P [i..j]$ of $P$ such that
\begin{itemize}
\item $P [i..j]$ occurs in $T$,
\item $i = 0$ or $P [i - 1..j]$ does not occur in $T$,
\item $j = m - 1$ or $P [i..j + 1]$ does not occur in $T$.
\end{itemize}
Finding long MEMs is an important task in bioinformatics and there are many tools for it.

Li~\cite{Li12} gave a practical algorithm, called forward-backward, for finding all the MEMs of $P$ with respect to $T$ using FM- or RLFM-indexes for $T$ and its reverse $T^\rev$.  Assume all the distinct characters in $P$ occur in $T$; otherwise, we split $P$ into maximal substrings consisting only of copies of characters occurring in $T$ and find the MEMs of those with respect to $T$.  We first use the index for $T^\rev$ to find the longest prefix $P [0..e_1]$ of $P$ that occurs in $T$, which is the leftmost MEM.  If $e_1 = m - 1$ then we are done; otherwise, $P [e_1 + 1]$ is in the next MEM, so we use the index for $T$ to find the longest suffix $P [s_2..e_1 + 1]$ of $P [0..e_1 + 1]$ that occurs in $T$.  The next MEM starts at $s_2$, so conceptually we recurse on $P [s_2..m - 1]$.  The total number of backward steps in the two indexes is proportional to the total length of all the MEMs.

Gagie~\cite{Gag24} proposed a heuristic for speeding up forward-backward when we are interested only in MEMs of length at least $L$.  We call this heuristic Boyer-Moore-Li, following a suggestion from Finlay Maguire~\cite{FM24}.  Since any MEM of length at least $L$ starting in $P [0..L - 1]$ includes $P [L - 1]$, we first use the index for $T$ to find the longest suffix $P [s..L - 1]$ of $P [0..L - 1]$ that occurs in $T$.  If $s = 0$ then we fall back on forward-backward to find the leftmost MEM and the starting position of the next MEM.  Otherwise, since we know there are no MEMs of length at least $L$ starting in $P [0..s - 1]$, conceptually we recurse on $P [s..m - 1]$.  Li~\cite{Li24} tested Boyer-Moore-Li and found it practical enough that he incorporated it into his tool {\tt ropebwt3}.

Suppose we build an {\tt rsync}-like parse of $T [0..n - 1]$ and two-level indexes for $T$ and $T^\rev$ based on that parse and parse $P$ when we get it.  With a na\"ive two-level version of Boyer-Moore-Li, we would simply use the two-level indexes in place of the normal FM- or RLFM-indexes for $T$ and $T^\rev$.  We conjecture, however, that we can do better in practice.

Let $P [k]$ be the last character of the last phrase that ends strictly before $P [L]$, let $P [j]$ be the first character of the first phrase such that $P [j..k]$ occurs in $T$, and let $P [i]$ be the second character of the phrase preceding the one containing $P [j]$.  Notice we can find $i$, $j$ and $k$ by matching phrase by phrase using only the top level (for the parse) of the two-level index for $T$.  If $i > 0$ then we can immediately discard $P [0..i - 1]$ and conceptually recurse on $P [i..m - 1]$; otherwise, we proceed normally.

Of course, the value $i$ is at most the value $s$ found by regular Boyer-Moore-Li and could be much smaller, in which case discarding $P [0..i - 1]$ benefits us much less than discarding $P [0..s - 1]$.  We hope this is usually not the case and we look forward to testing Boyer-Moore-Li with two-level indexing.

\end{document}